\documentclass[aps,preprint]{revtex4}
\usepackage{graphics,epsfig,amssymb}

\usepackage{subfigure}

\def\beq{\begin{equation}} 
\def\eeq{\end{equation}} 
\begin{document}

\title{Coexistence of quartets and pairs in even-even $N>Z$ nuclei}

\author{M. Sambataro$^{a)}$, N. Sandulescu$^{b)}$ and D. Gambacurta$^{c)}$}
\affiliation{$^{a)}$Istituto Nazionale di Fisica Nucleare - Sezione di Catania,
Via S. Sofia 64, I-95123 Catania, Italy \\
$^{b)}$National Institute of Physics and Nuclear Engineering, P.O. Box MG-6, 
Magurele, Bucharest, Romania\\
$^{c)}$INFN-LNS, Labotatori Nazionali del Sud, 95123 Catania, Italy}

\begin{abstract}
We analyse the structure of the ground states of  even-even $N>Z$ nuclei with nucleons
moving in the same major shell and interacting via realistic two-body forces
of shell-model type. We express the ground states of these nuclei as a product of a 
quartet term, which represents the $N = Z$ subsystem,
and a pair condensate built with  the excess neutrons. The accuracy of this approximation
is discussed for nuclei with valence nucleons in the $sd$ and $pf$ major shells. 
\end{abstract}

\maketitle

\section{Introduction}

A specific feature of $N=Z$ nuclei is the  occurrence of quartet structures, composed by two neutrons
and two protons, which have strong internal correlations and interact weakly with each other. 
If these 4-body structures are well localised in space, they are usually referred to 
as $\alpha$-clusters. The well known example is the  $\alpha$-clustering in light $N=Z$ nuclei,
which has been predicted since 30'ties 
\cite{bethe,wefelmeier,wheeler,hafstad} and systematically studied
afterwards in the framework of $\alpha$-cluster models \cite{morinaga,brink,ikeda,arima,freer} 
(for more recent studies, see \cite{otsuka} and  references therein).

In the 1960's it was pointed out that in the ground state of $N=Z$ nuclei a more  general type of
quartet structures can appear which are  induced by the proton-neutron pairing interaction.
In this case the 4-body correlations manifest in the configuration space rather 
than in the real space \cite{soloviev,flowers,valatin}. These correlations bear
resemblance to the pairing correlations between like-particles and they were represented
initially by a BCS-like state expressed in terms of quartets \cite{flowers,valatin}. The idea that 
quartets provided an appropriate tool to describe the ground states of proton-neutron 
pairing Hamiltonians in $N=Z$ nuclei
was explored later on in various studies \cite{eichler,dobes,senkov,chasman}. 
An important step forward was the finding that the ground state of these  systems can be well 
described by a very simple state: a condensate of $\alpha$-like quartets \cite{eichler,dobes,qcm_t1,qcm_t0t1,qm_qcm_t0t1}.
Moreover, it was found that the ground states of proton-neutron pairing Hamiltonians for $N>Z$ systems have also a simple structure: 
a condensate of quartets to which it is attached a pair condensate formed by the excess neutrons \cite{dobes,qcm_t1_ngz,qcm_t0t1_ngz}. 
The scope of this study is to investigate to what extent similar quartet-pair structures  can be
identified in  realistic calculations of  $N>Z$ nuclei with nucleons moving
in the same major shell and interacting via general two-body forces of shell-model type.

The role of quartets in the framework of the shell-model has been investigated quite in detail for $N=Z$ nuclei \cite{arima_gillet,hasegawa,qm_prl,qm_pd,qm_odd,qm_epja,qm_ex}. The same
cannot be said for the case of $N>Z$ nuclei. The approach which appears as the most appropriate to describe the
coexistence of quartets and pairs in the shell model framework is the one proposed many years ago by Arima and Gillet \cite{arima_gillet}. In this approach the eigenstates 
are calculated in a quartet-pair basis formed by products of  quartets and neutron pairs of various
angular momenta. This calculation scheme was applied to a schematic system composed by 4 protons and 4 neutrons 
sitting on two different shells. In the present article  we propose a simpler quartet-pair approach,
adapted for the the case of nucleons moving in the same valence shell and inspired by the 
proton-neutron pairing models mentioned above \cite{qcm_t1_ngz,qcm_t0t1_ngz}. Namely, we suppose 
that the ground states of $N>Z$ nuclei can be approximated by a product between a quartet core,
which  represents  the $N=Z$ subsystem, and a pair condensate built with the 
extra neutrons. The calculation follows an iterative scheme where, alternatively, the quartet core
and the pairs are fixed through a minimisation procedure. This approximation scheme will be applied
to $N>Z$ nuclei with valence nucleons moving in the $sd$ and $pf$ major shells and interacting
via realistic shell-model interactions. The comparison of exact and approximate energies and
occupation numbers will be employed to judge the quality of the present quartet-pair approximation.

The paper is organized as follows. In Section II, we will describe the formalism and show the results. In Section III, we will draw the conclusions. 

\section{Procedure and results}

This work shall focus on the analysis of ground states of even-even nuclei with $N>Z$ and both protons and neutrons occupying the same orbits.  
These ground states will be approximated as a product of a core formed by $n$ isospin $T=0$ quartets and a condensate of $m=(N-Z)/2$ pairs.
By labeling as $N_\pi$ and $N_\nu$ the number of protons and neutrons which are outside the closed shell, the number of quartets in the core is $n=N_\pi /2$.

We assume a spherically symmetric mean field and, using the standard notation, we introduce the label
$i\equiv \{n_i,l_i,j_i\}$ to identify the orbital quantum numbers. We define the  $T=0$ quartet creation operator as
\begin{equation}
q^+_{JM}=\sum_{i_1j_1J_1}\sum_{i_2j_2J_2}\sum_{T'}
q_{i_1j_1J_1,i_2j_2J_2,{T'}}
[[a^+_{i_1}a^+_{j_1}]^{J_1{T'}}[a^+_{i_2}a^+_{j_2}]^{J_2{T'}}]^{JT=0}_{M},
\label{1}
\end{equation}
where $a^+_i$ creates a fermion on the orbital $i$ and $M$ stands for the projection of $J$.
No restrictions on the intermediate couplings $J_1T'$ and $J_2T'$ are introduced.

In order to fix the quartet-pair approximation of the ground state
we proceed through an iterative procedure. As an initial step, we search for an approximation of the quartet core. To this purpose, following the scheme of Ref. \cite{sasa_band}, we introduce the state
\beq
|\Theta_{n}\rangle = (Q^+)^n|0\rangle ,
\label{2}
\eeq
where
\beq
Q^+ =\sum_J q^+_{J0}
\label{3}
\eeq
and $n$ is number of quartets which characterizes the $N=Z$ core.
$|\Theta_{n}\rangle$ is thus a condensate of $n$ quartets $Q^+$, each of these quartets being in turn  a linear superposition of the quartets $q^+_{J0}$ (\ref{1}) whose angular momentum $J$ runs over a set of values to be specified. 
By minimizing the energy of the state $|\Theta_{n}\rangle$ we fix the quartets $q^+_{J}$. The condensate $|\Theta_{n}\rangle$ has a total isospin $T=0$ but no well defined angular momentum. In order to construct the $J=0$ ground state
of the quartet core, we carry out
a configuration-interaction calculation in a space spanned by the quartets $q^+_{J}$. To do so we define the set of states (we work in the $m$-scheme)
\begin{equation}
|\Lambda^{(n)}_{\overline M},\{N_{JM}\}\rangle = \prod_{J\in{(0,J_{max})}; M\in{(-J,J)} }(q^+_{JM})^{N_{JM}}|0\rangle
\label{1a}
\end{equation}
with the conditions
\beq
\sum_{JM}N_{JM}=n,~~~~~~~\sum_{JM}MN_{JM}=\overline{M}.
\eeq
We then orthonormalize the states (\ref{1a}) and diagonalize the Hamiltonian in this new basis for the various $\overline{M}$. The lowest eigenstate, $|\Psi_0(n)\rangle$, provides the initial $J=0, T=0$ approximation for the $N=Z$ core.

Having fixed $|\Psi_0(n)\rangle$, as a second step, we  introduce the neutron condensate. This is done by defining the state 
\beq
|\Omega (m,n)\rangle =(P^+)^m|\Psi_0(n)\rangle,
\label{4}
\eeq
 where
\beq
P^+=\sum_i\alpha_i[a^+_ia^+_i]^{J=0,T=1}_{M_T=1}
\label{5}
\eeq
creates a collective neutron pair with $J=0$. By minimizing the energy of the state $|\Omega (m,n)\rangle$ with respect to the pair amplitudes $\alpha_i$ we fix the neutron condensate. 
This defines the first-order approximation of the ground state of the $N>Z$ system. The iterative procedure proceeds  by minimizing the energy of the state $|\Omega (m,n)\rangle$ with
respect to variations, alternatively, of $|\Psi_0(n)\rangle$ (keeping
fixed the neutron condensate and the structure of the quartets $q^+_{J}$) and of the neutron condensate (keeping fixed $|\Psi_0(n)\rangle$).
In all the calculations presented below we have restricted the values of the angular momentum of the quartets $q^+_{J}$ to $J=0,2,4$.

\begin{figure*}[h]
\centering
\mbox{\subfigure{\includegraphics[scale=0.3, angle=-90]{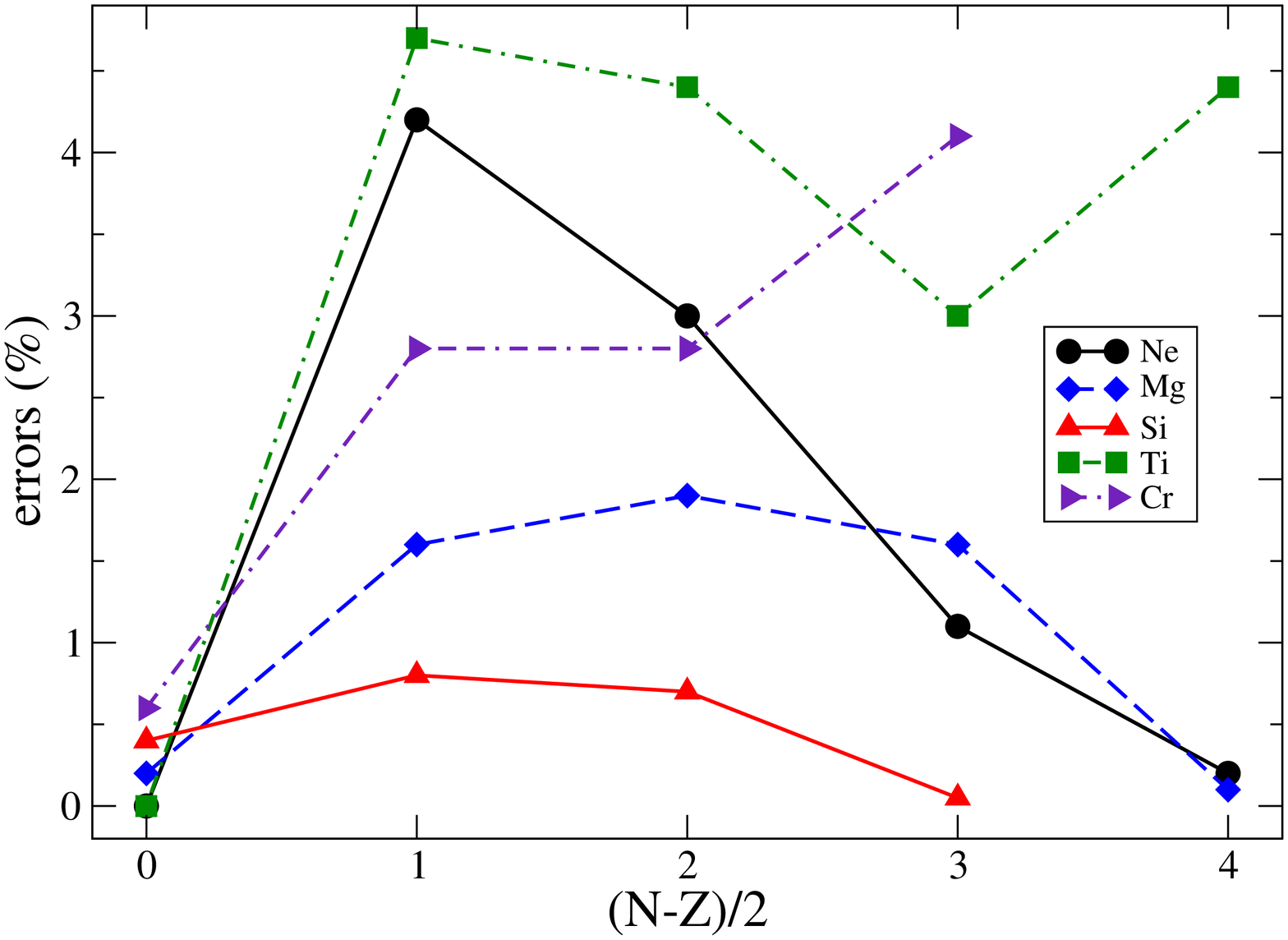}}\quad
\subfigure{\includegraphics[scale=0.3, angle=-90]{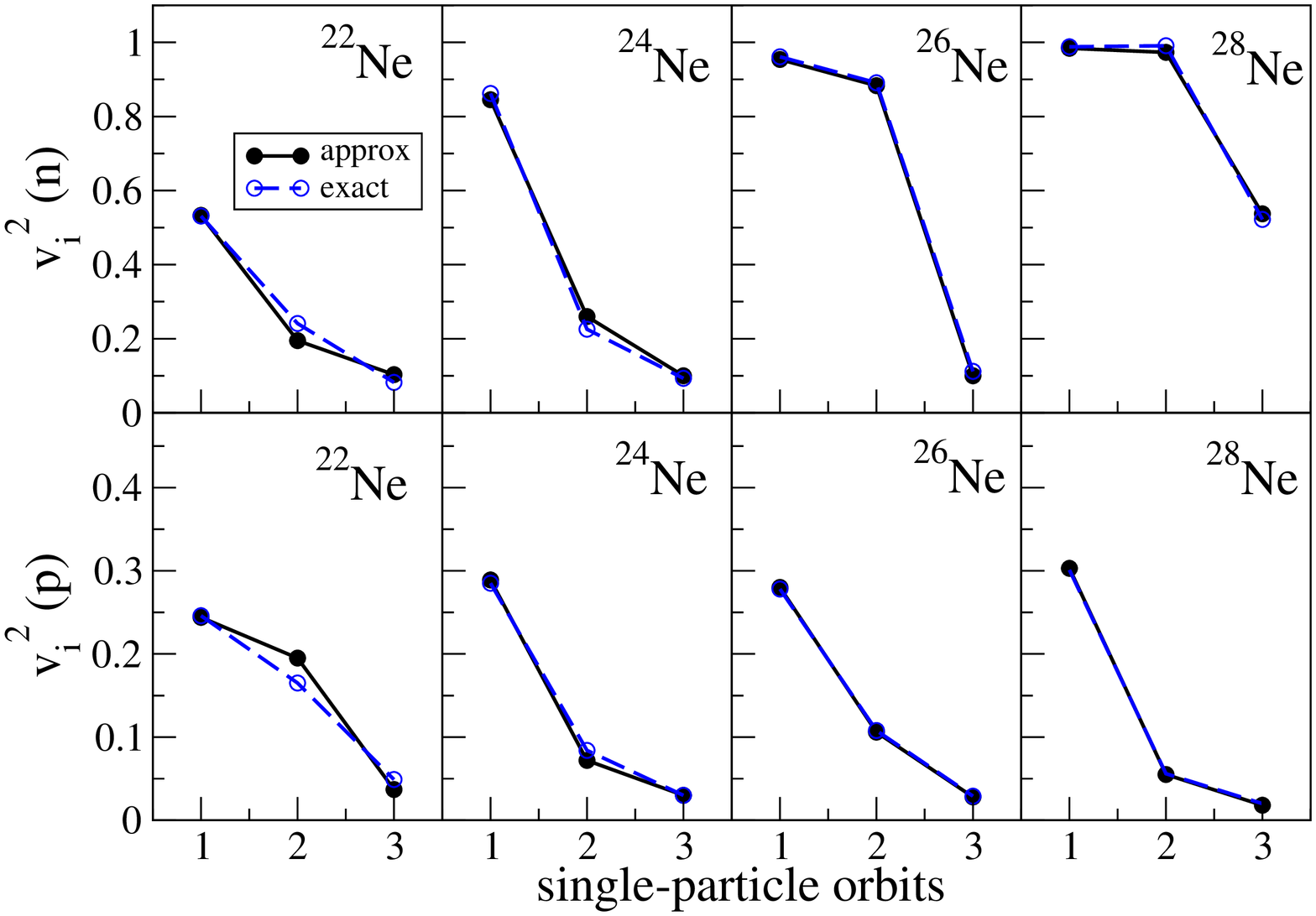}}}
\caption{(Left) Errors for the ground state energies predicted by the approximation (6). 
(Right) Occupation probabilities of the orbits $1d_{5/2}$ (1), $2s_{1/2}$ (2) and $1d_{3/2}$ (3) 
for  Ne isotopes.}
\end{figure*}

To test the approximation (6) we take as example nuclei with the valence nucleons in the sd and pf shells, 
which are described with the realistic two-body interactions  USDB \cite{usdb} and, respectively, KB3G \cite{kb3g}.
We consider various isotopic chains, starting with the N=Z nucleus to which we add, progressively, 
neutron pairs. The ground states of these nuclei are evaluated 
with the approximation (6) and the results are  contrasted with the exact shell-model calculations
obtained with the code BIGSTICK \cite{bigstick}.
The accuracy of  the approximation (6) for the ground state energies is illustrated in Fig.1, left panel. 
For each isotopic chain we show the relative errors in the ground state energies.
First of all it can be noticed that
the errors for the self-conjugate nuclei
with 2 and 3 quartets outside the closed shells are very small, under $0.7\%$. By adding a neutron pair to
the self-conjugate nuclei the errors increase significantly. This is especially the case for Ne and
Ti isotopes. For these isotopes it can also be seen that the errors fall rapidly when we pass from
one to two extra neutron pairs. In general, by adding more than two neutron pairs, the errors decrease in 
all the $sd$ nuclei, which reflects the filling of the second half of the major shell.  
A similar behaviour can be noticed for the Ti isotopes due to the filling of the $f_{7/2}$ orbit. 
As it can be seen, after this orbit is filled, the errors increase again when one more pair is 
added to Ti and Cr isotopes. Overall, as expected, the errors are the biggest when the nucleons have 
a larger phase space for building up correlations. However, in all analysed nuclei the errors
remain small, below $5\%$, which shows that the approximation (6) is working reasonably well.

\begin{figure*}[h]
\centering
\mbox{\subfigure{\includegraphics[scale=0.3, angle=-90]{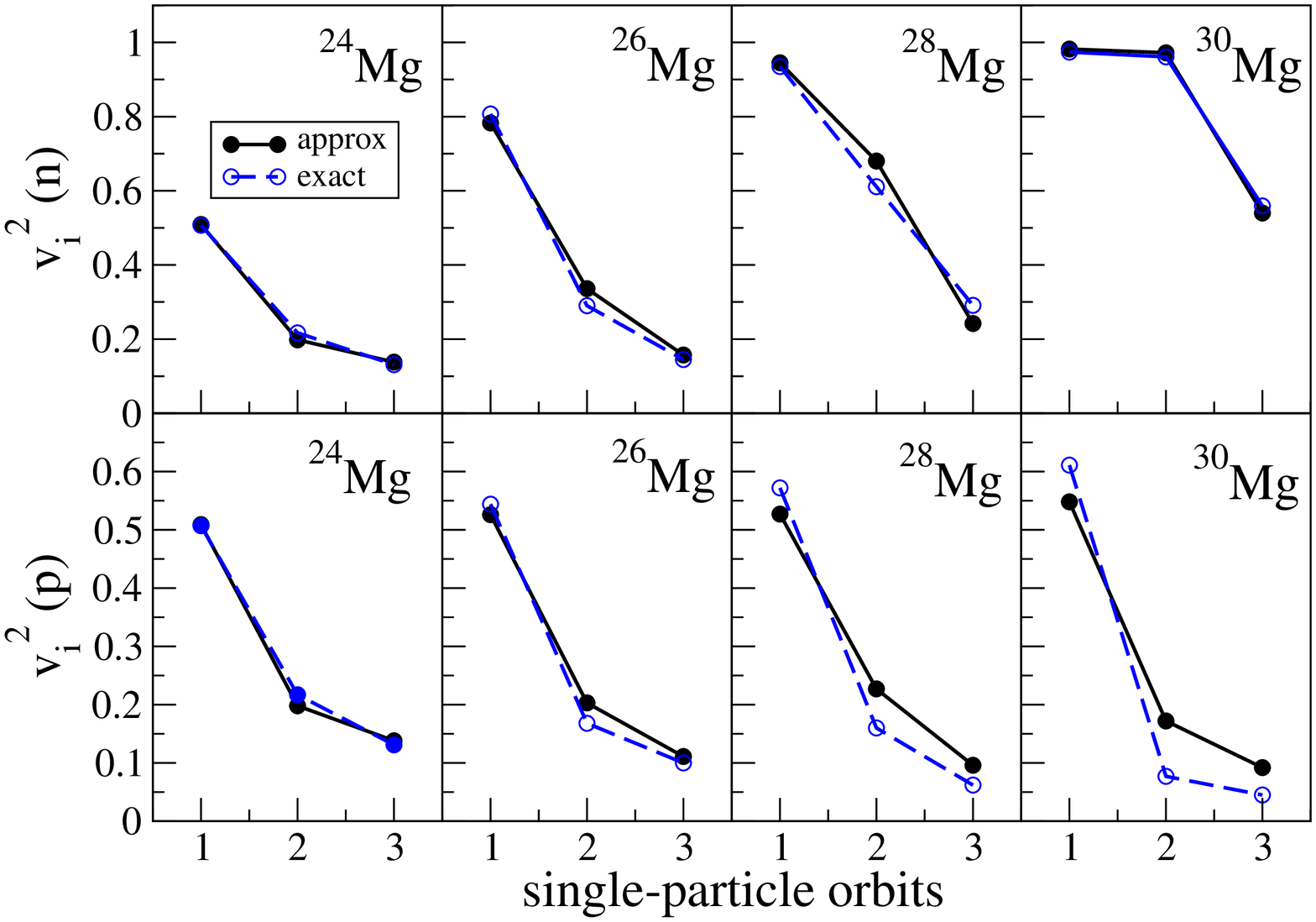}}\quad
\subfigure{\includegraphics[scale=0.3, angle=-90]{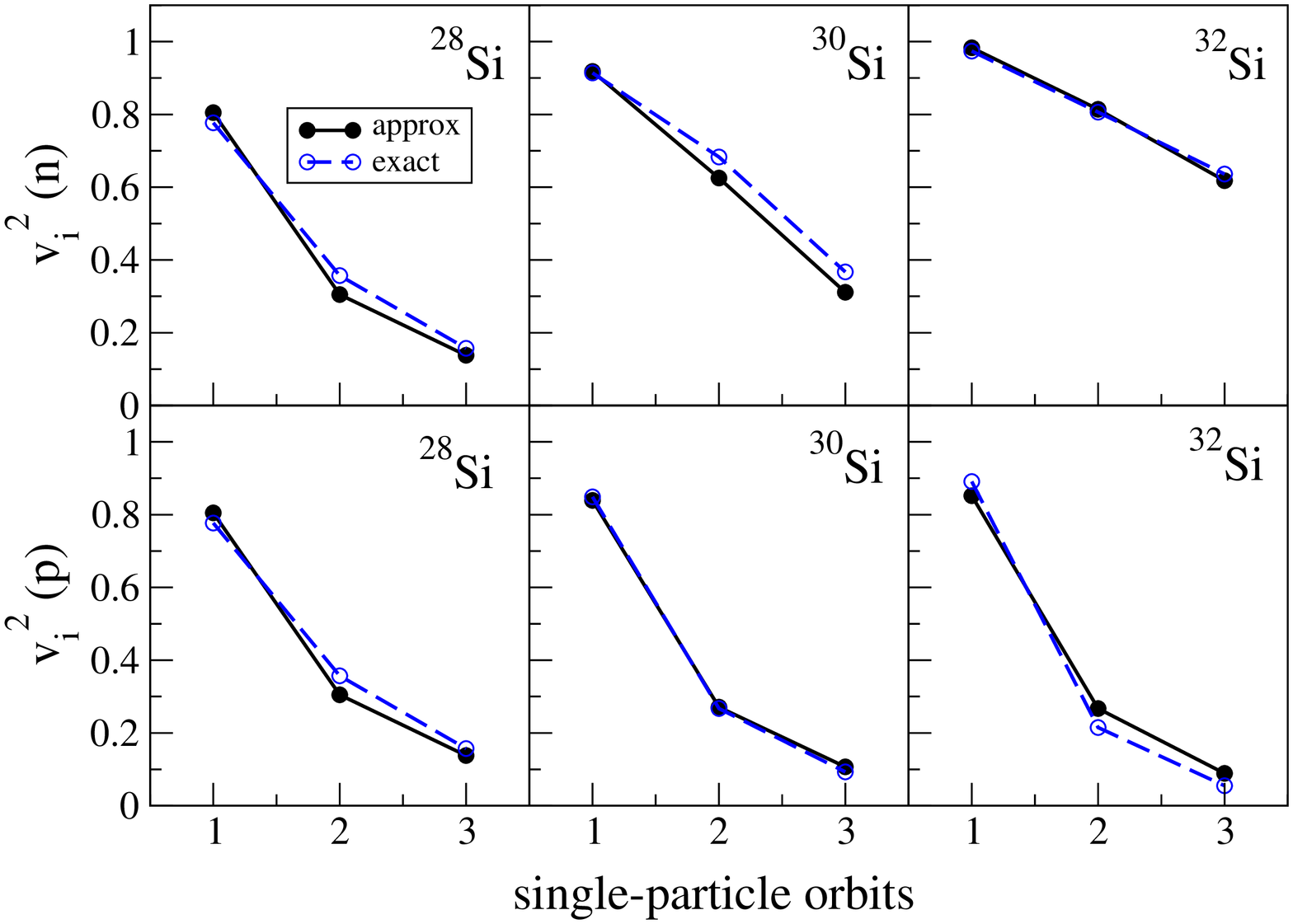}}}
\caption{Occupation probabilities of the orbits $1d_{5/2}$ (1), $2s_{1/2}$ (2) and $1d_{3/2}$ (3) 
for Mg and Si isotopes.}
\end{figure*}

\begin{figure*}[h]
\centering
\mbox{\subfigure{\includegraphics[scale=0.3, angle=-90]{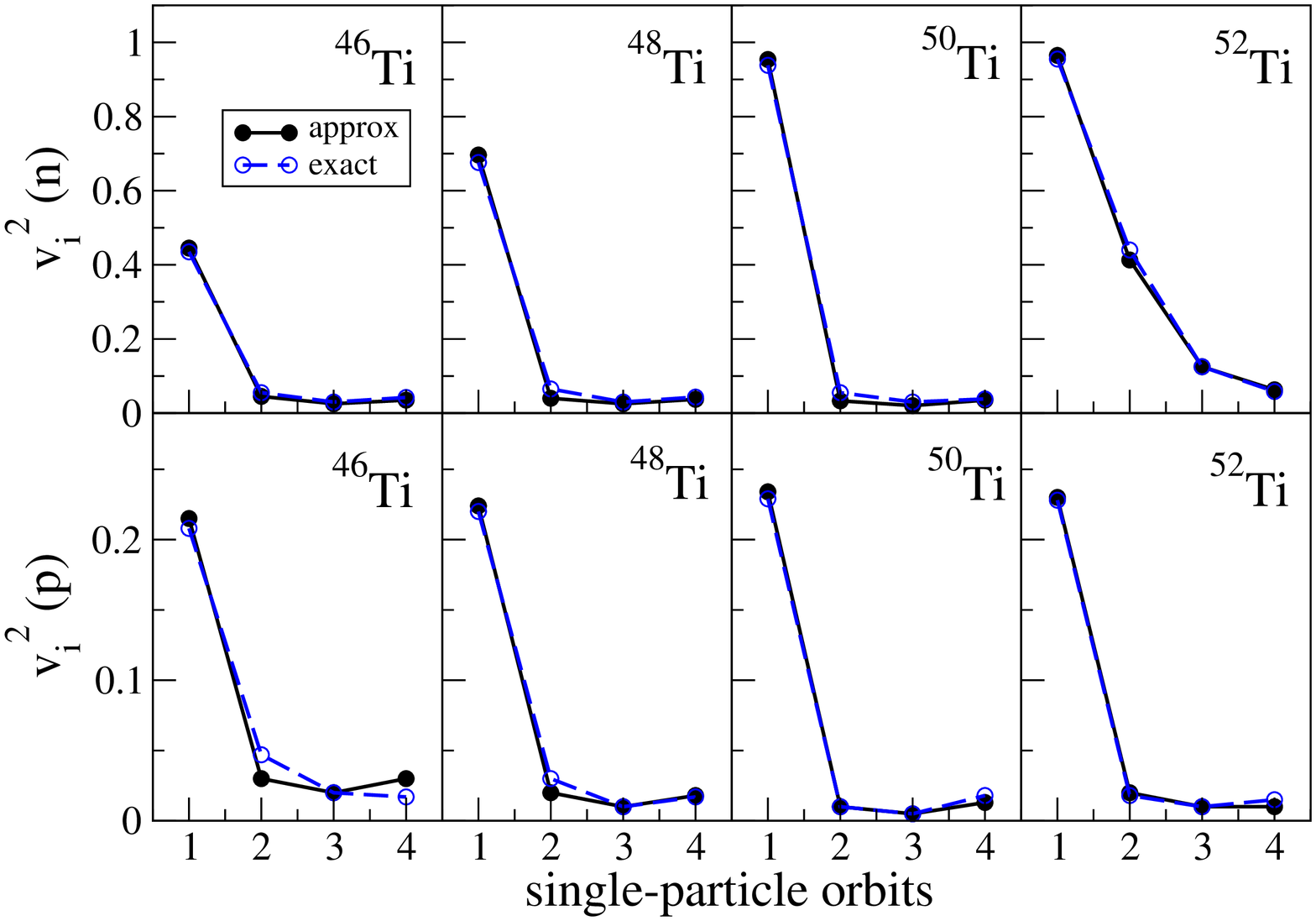}}\quad
\subfigure{\includegraphics[scale=0.3, angle=-90]{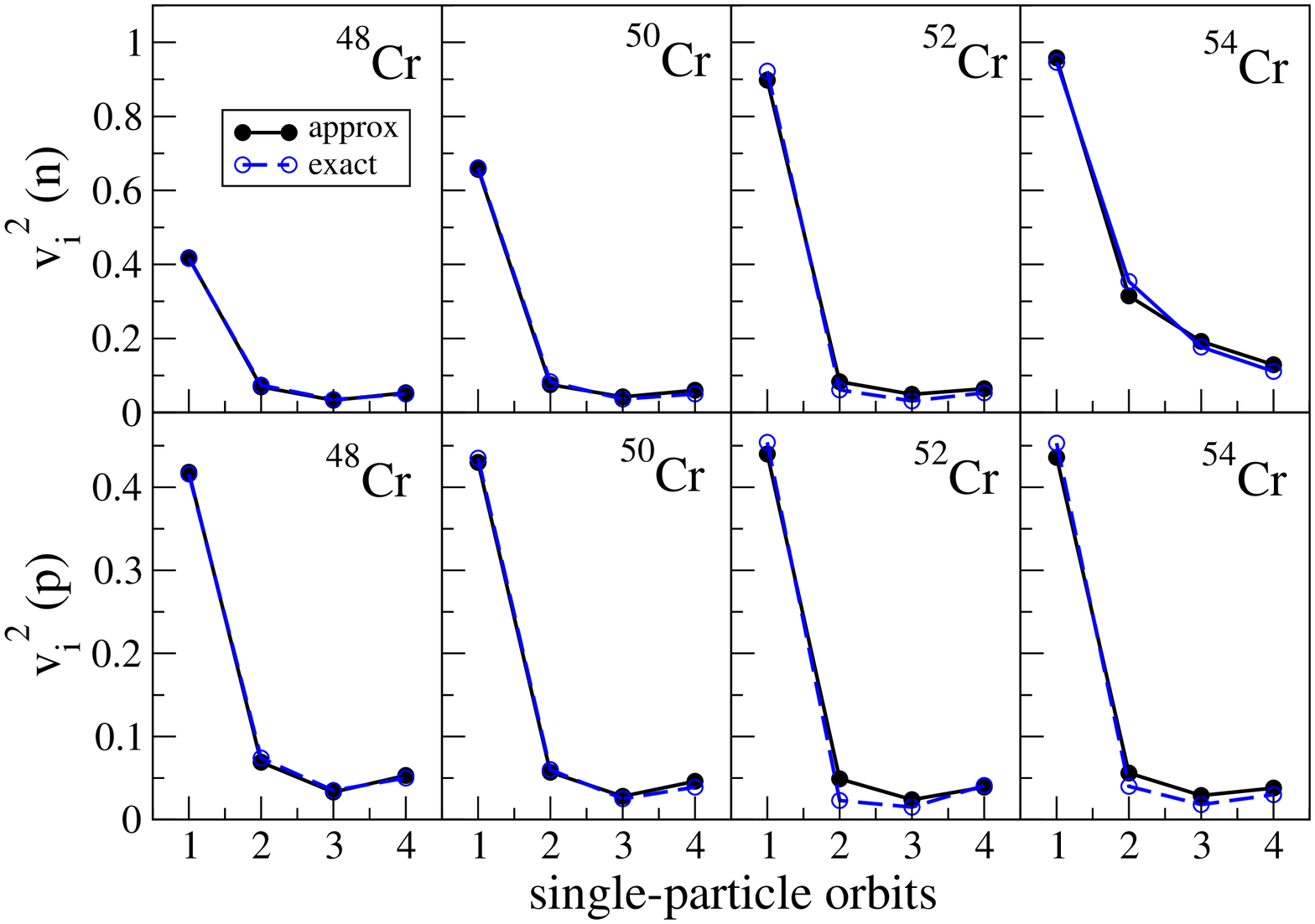}}}
\caption{Occupation probabilities of the orbits $1f_{7/2}$ (1), $2p_{3/2}$ (2), $2p_{1/2}$ (3) and $1f_{5/2}$ (4)
for Ti and Cr isotopes.}
\end{figure*}

\begin{figure*}[h]
\centering
\mbox{\subfigure{\includegraphics[scale=0.3, angle=-90]{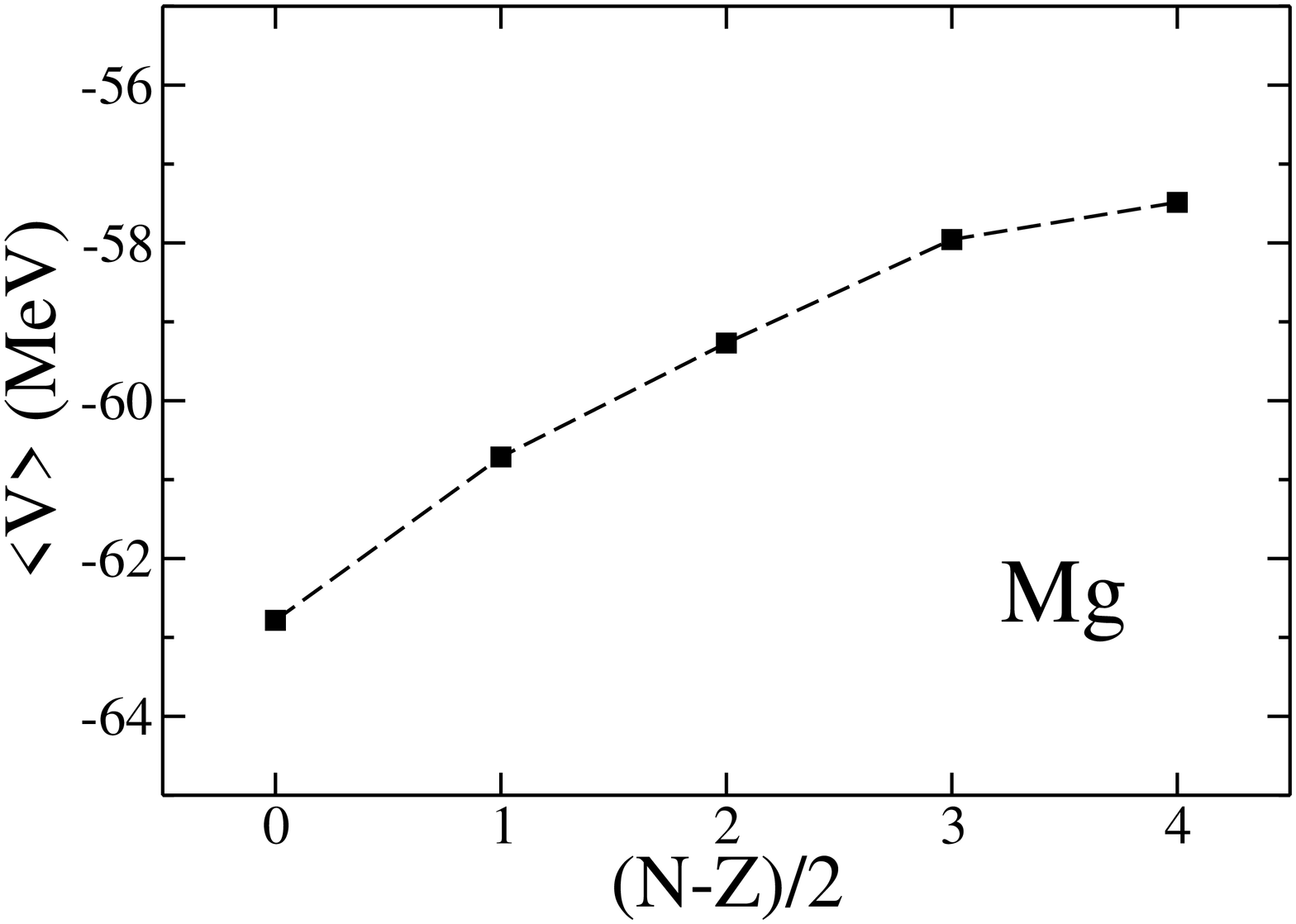}}\quad
\subfigure{\includegraphics[scale=0.3, angle=-90]{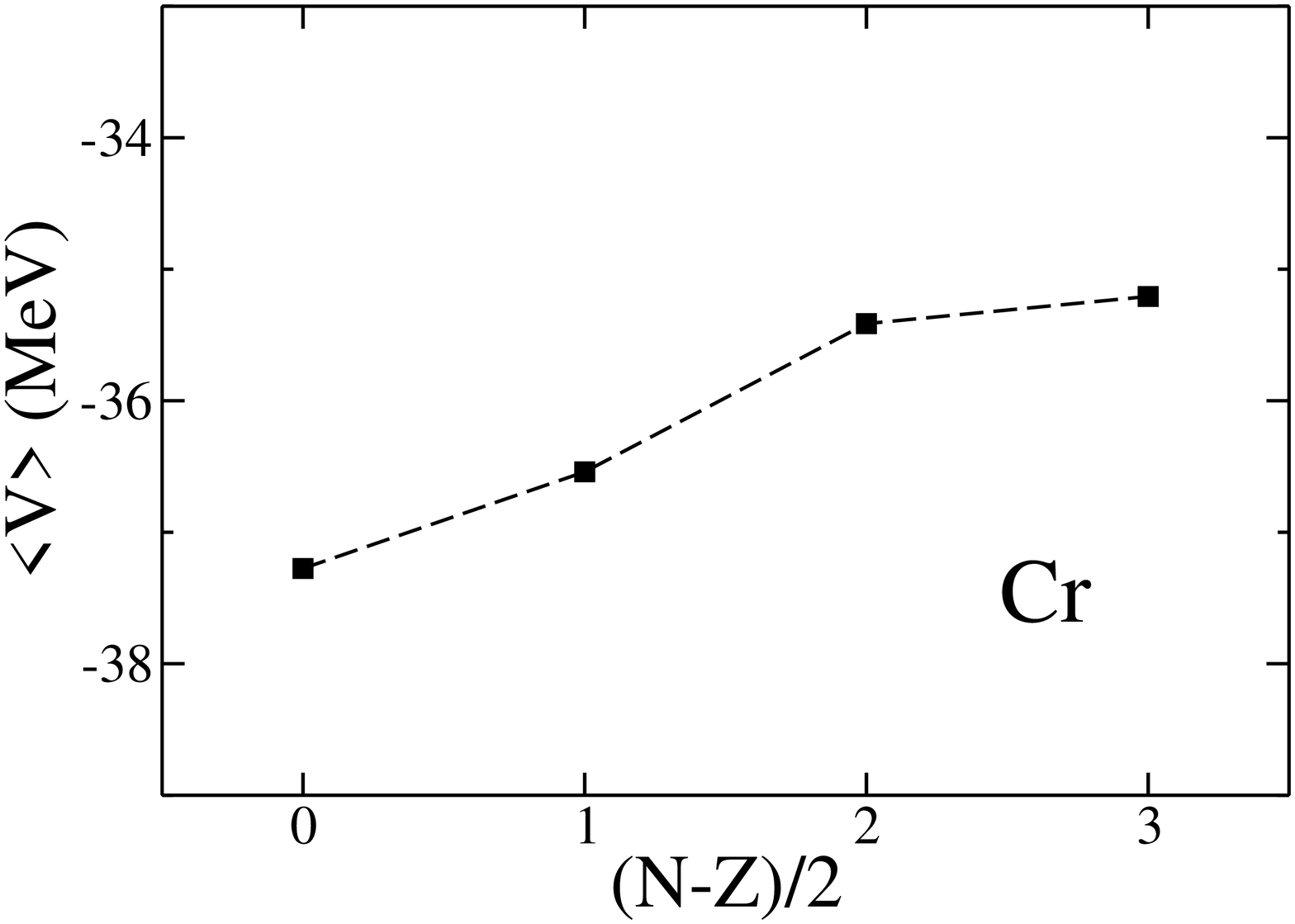}}}
\caption{The quantity $\langle V \rangle $ (8) as a function of neutron excess for
Mg (left) and Cr (right) isotopes.}
\end{figure*}

In Fig. 1, right panel,  and Figs. 2-3 we show the occupation probabilities of the single-particle orbits of protons 
and neutrons predicted by the approximation (6). It can be seen that the predictions follow fairly
well the exact results obtained by diagonalisation. It is worth noticing how the distribution of the protons 
changes when neutrons are added to the self-conjugate isotope. As expected, the occupancy of the lowest orbit, 
$d_{5/2}$ for the $sd$ nuclei and $f_{7/2}$ for the $pf$ nuclei,  increases with increasing the
neutron excess. This is mainly due to the decrease of correlations in the $N=Z$
subsystem which follows the addition of extra neutrons. As seen in Fig. 2, this effect is overestimated 
by the approximation (6) in the case of the heaviest Mg isotopes.

An indication of how much the extra neutrons affect the quartet correlations in the $N=Z$ subsystem
can be obtained from the quantity 
\begin{equation}
\langle V \rangle = \langle \Psi_0 (n) |V| \Psi_0 (n) \rangle / \langle \Psi_0 (n)| \Psi_0 (n) \rangle
\end{equation}
where V is the two-body interaction and $|\Psi_0 (n) \rangle$ is the wave function of the $N=Z$ quartet core.
As an example, in Fig. 4 we show the 
dependence of $ \langle V \rangle $ on the number of extra neutrons for Mg and Cr isotopes. It can be seen that
the extra neutrons suppress significantly the quartet correlations in the $N=Z$ subsystem. A similar behaviour
was noticed before in the calculations done with the proton-neutron pairing interactions 
\cite{qcm_t1_ngz,qcm_t0t1_ngz,virgil}. 

\section{Summary and conclusions}

The ground states of even-even $N>Z$ nuclei, with neutrons and protons moving in the same major shell, 
have been  described by a product of two terms, one representing the $N=Z$ subsystem and the other one the
excess neutrons.  The first term is expressed by quartets built of two neutrons and two protons
coupled to total isospin $T=0$. The second term is represented as a neutron condensate.
The structure of the quartet core and of the neutron pairs have been determined variationally from the minimisation of the
energy.  This approach has been applied to nuclei with valence nucleons in the $sd$ and $pf$ major shells.
It has been shown that the quartet-pair approximation scheme provides ground state energies which are in fair agreement
with the exact shell-model values. In addition, the occupation probabilities of the single-particle orbits 
predicted by the quartet-pair model follow reasonably well the exact probabilities. 
These results indicate
that the ground state of even-even $N>Z$ nuclei can be represented to a good extent in a simple form,
which resembles that adopted for systems interacting via proton-neutron pairing  forces \cite{qcm_t1_ngz,qcm_t0t1_ngz}. Namely, as a product between a quartet state representing 
the N = Z subsystem and a condensate of pairs formed with the extra neutrons.
  
\vskip 0.3cm
{\it Acknowledgments}\\ 
This work was supported by a grant of the Romanian Ministry of Research and Innovation, CNCS - UEFISCDI, 
project number PCE 160/2021, within PNCDI III.


\begin{thebibliography}{10}
\bibitem{bethe}
H.A. Bethe, R.F. Bacher, Rev. Mod. Phys. {\bf 8}, 82 (1936).
\bibitem{wefelmeier}
W. V. Wefelmeier,  Z. Phys. Hadrons Nucl. {\bf 107}, 332 (1937).
\bibitem{wheeler}
J. A.  Wheeler,  Phys. Rev. {\bf 52}, 1083 (1937).
\bibitem{hafstad}
L.R. Hafstad, E. Teller, Phys. Rev. {\bf 54}, 681 (1938).
\bibitem{morinaga}
3. H. Morinaga, Phys. Rev. C. {\bf 101}, 254 (1956).
\bibitem{brink}
D. Brink, Proc. Int. Sch. Phys. Enrico Fermi. Course {\bf 36}, 247 (1966).
\bibitem{ikeda}
Ikeda, K., Takigawa, and H. Horiuchi, Prog. Theor. Phys. Suppl. E{\bf 68}, 464 (1968).
\bibitem{arima}
 A. Arima, H. Horiuchi, K. Kubodera, N. Takigawa, Clustering in Light Nuclei. 
In: Baranger M. and Vogt E. (ed) Advances in Nuclear Physics, 5, 345 (Springer, 
Boston, MA, 1973).
\bibitem{freer}
M. Freer, H. Horiuchi, Y. Kanada-En’yo, D. Lee, U. Meißner,  Rev. Mod. Phys. {\bf 90}, 035004 (2018).
\bibitem{otsuka} 
T. Otsuka, T. Abe, T. Yoshida, Y. Tsunoda, N. Shimizu, N. Itagaki, Y. Utsuno,
J. Vary, P. Maris, H. Ueno, Nature Communications {\bf 13}, 2234 (2022). 
\bibitem{soloviev} 
V. G. Soloviev, Nucl. Phys. {\bf 18}, 161 (1960).
\bibitem{flowers}
B.H. Flowers and M. Vujicic, Nucl. Phys. {\bf 49}, 586 (1963).
\bibitem{valatin} 
B. Bremond and J. G. Valatin, Nuclear Physics {\bf 41} (1963) 640.
\bibitem{eichler}
 J. Eichler and M. Yamamura, Nucl. Phys. A {\bf 182}, 33 (1972).
\bibitem{dobes}
J. Dobes and S. Pittel, Phys. Rev. C {\bf 57}, 688 (1998).
\bibitem{senkov}
R. A. Senkov and V. Zelevinsky, Phys. At. Nucl. {\bf 74}, 1267
(2011).
\bibitem{chasman} R. R. Chasman, Phys. Lett. B {\bf 524}, 81 (2002).
\bibitem{qcm_t1}
 N. Sandulescu, D. Negrea, J. Dukelsky, C.W. Johnson, Phys. Rev. C 
{\bf 85}, 061303(R) (2012).
\bibitem{qcm_t0t1} 
N. Sandulescu, D. Negrea, and D. Gambacurta, Phys. Lett. B {\bf 751}, 348 (2015).
\bibitem{qm_qcm_t0t1} 
M. Sambataro and N. Sandulescu, Phys. Rev. C {\bf 93}, 054320 (2016).
\bibitem{qcm_t1_ngz}
 N. Sandulescu, D. Negrea, C. W. Johson, Phys. Rev. C {\bf 86}, 041302(R) (2012). 
\bibitem{qcm_t0t1_ngz} 
D. Negrea, P. Buganu, D. Gambacurta, N. Sandulescu, Phys. Rev. C {\bf 98}, 064319 (2018).
\bibitem{virgil}
V.V. Baran and D.S. Delion, Phys. Rev. C {\bf 100}, 034326 (2019).
\bibitem{arima_gillet}
A. Arima and V. Gillet, Annals of Physics {\bf 66}, 117 (1971).
\bibitem{hasegawa}
M. Hasegawa, S. Tazaki, and R. Okamoto, Nucl. Phys. A {\bf 592}, 45 (1995).
\bibitem{qm_prl}
M. Sambataro and N. Sandulescu, Phys. Rev. Lett. {\bf 115}, 112501 (2015).
\bibitem{qm_pd}
M. Sambataro and N. Sandulescu, Phys. Rev. C {\bf 91}, 064318 (2015).
\bibitem{qm_odd}
M. Sambataro and N. Sandulescu, Phys. Lett. B {\bf 763}, 151 (2016).
\bibitem{qm_epja} M. Sambataro and N. Sandulescu, Eur. Phys. J. A {\bf 53}, 47 (2017).
\bibitem{qm_ex} M. Sambataro and N. Sandulescu, Phys. Lett.  B {\bf 820}, 136476 (2021).
\bibitem{usdb}B.A. Brown and W.A. Richter, Phys. Rev. C {\bf 74}, 034315 (2006).
\bibitem{kb3g} A. Poves and G. Martinez-Pinedo, Phys. Lett B {\bf 430}, 203 (1998).
\bibitem{sasa_band}
M. Sambataro and N. Sandulescu, Phys. Lett. B {\bf 827}, 136987 (2022). 
\bibitem{bigstick} C. W. Johnson, W. E. Ormand, P. G. Krastev, Comp. Phys. Comm. {\bf 184}, 2761 (2013).
\end{thebibliography}
\end{document}